\documentstyle[12pt]{article}
\newtheorem{result}{Result}
\newtheorem{theorem}{Theorem}
\newtheorem{lemmas}{Lemma}
\newtheorem{conj}{Conjecture}
\begin{document}
\parindent 8pt
\begin{titlepage}
\rightline{IASSNS-HEP-94/22}\par
\begin{center}
{\Large \bf Localized Exotic Smoothness}
\end{center}\par
\medskip
\begin{center}
{\bf Carl H. Brans}
\end{center}\par
\medskip
\begin{center}

{Institute for Advanced Study\\ Princeton, NJ 08540\\ and\\
Physics Department\\ Loyola University\\ New Orleans, LA 70118  \\
e-mail:brans@music.loyno.edu}\par\bigskip
March 30, 1994
\end{center}
\begin{abstract}
Gompf's end-sum techniques are used to establish the existence of an
infinity of non-diffeomorphic manifolds, all having the same trivial
${\bf R^4}$ topology, but for which the exotic differentiable structure
is confined to a region which is spatially limited.  Thus, the smoothness
is standard outside of a region which is topologically (but not smoothly)
${\bf B^3}\times {\bf R^1}$, where ${\bf B^3}$ is the compact three ball.
The exterior of this region is diffeomorphic to standard
${\bf R^1}\times {\bf S^2}\times{\bf R^1}$.  In a space-time diagram, the
confined exoticness sweeps out a world tube which, it is conjectured,
might  act as a source for certain non-standard solutions to the Einstein
equations.  It is shown that smooth Lorentz signature metrics can be
globally continued from ones given on appropriately defined regions,
including the exterior (standard) region. Similar constructs are provided
for the topology,
${\bf S^2}\times {\bf R^2}$
of the Kruskal form of the Schwarzschild solution.  This leads to
conjectures on the existence of Einstein metrics  which are externally
identical to standard black hole ones, but none of which can be globally
diffeomorphic to such standard objects. Certain aspects of the Cauchy
problem are also discussed in terms of
${\bf R^4_\Theta}$\
models which are ``half-standard'', say for all $t<0,$ but for which
$t$ cannot be globally smooth.\end{abstract}
\end{titlepage}
This paper is concerned with smooth manifold models for space-time which
have relatively trivial topology, e.g., ${\bf R^4}$, or ${\bf R^2\times
S^2}$, but non-standard, or ``exotic'' smoothness structures.  By
definition, such manifolds are not diffeomorphic to their standard smooth
form, and hence, from the basic principles of general relativity, cannot
be physically equivalent to any previously studied manifold with the
corresponding simple topology.    In the non-compact cases an important
feature of many of these examples is that they  require that the exotic
part extend ``to infinity'', as illustrated, for example, in figure 1.
This fact has served as a deterrent to the consideration of such spaces
as space-time models since classical observations of space-time are
``large scale'' in some sense, with resulting expectation of asymptotic
regularity, including smoothness.
In fact, all that the mathematics requires is that the exotic region not
be contained in a compact set.
However, to my knowledge,  no example of an exotic manifold which is {\em
standard} at {\em spatial} infinity has ever been published before now.
The main result of this paper can be summarized informally: \par
\begin{result}
There exists exotic smooth manifolds with ${\bf R^4}$ topology which are
standard at spatial infinity, so that the exoticness can be regarded as
spatially confined.\end{result}\par
A more precise statement of this result is provided in Theorem 1 below.
The resulting manifold structures are illustrated in  examples such as
those shown in figures 3 and 4 where everything looks normal at space-
like infinity but the standard structure cannot be continued all the way
in to spatial origin. This work is based on the remarkable mathematical
breakthroughs of Milnor, Freedman, Donaldson, Gompf
\cite{miln},\cite{f},\cite{d},\cite{g0}, and others, establishing the
surprising existence of such exotic structures on topologically trivial
spaces, including ${\bf R^4}$, together with the end-sum techniques of
Gompf\cite{g1}.\par
This result could have great significance in all fields of physics, not
just relativity. Some model of space-time underlies every field of
physics.  It has now been proven that we cannot infer that space  is
necessarily smoothly standard from investigating what happens at space-
like infinity, even for topologically trivial
${\bf R^4}$. It seems very clear that this is potentially very important
to all of physics since it implies that there is another possible
obstruction,  in addition to material sources and topological ones, to
continuing external vacuum solutions for any field equations from
infinity to the origin.  Of course, in the absence of any explicit
coordinate patch presentation, no example can be displayed. However,
this\  leads naturally to a conjecture, informally stated: \par
\begin{conj}
This localized exoticness can act as a source for some externally regular
field, just as matter or a wormhole can.
\end{conj}\par
Of course, the exploration of this conjecture will require more detailed
knowledge of the global metric structure than is available at present.
The notions of domains of dependence, Cauchy surfaces, etc., necessary
for such studies cannot be fully explored with present differential
geometric information on exotic manifolds.  However, a beginning can be
made with certain general existence results as established and discussed
below.
\par
In order to provide some background for these matters, let us begin
with a brief review of the relevant mathematical facts.  The
apparently innocuous question of whether or not  the set of
differentiable
structures (modulo diffeomorphisms)
on ${\bf R^n}$ is trivial has long been of mathematical interest.  As
of about ten years ago, this question had been settled in the expected
affirmative for all $n\ne 4$, and probably most people expected the
exceptional case $n=4$ to ultimately resolve to the same conclusion.
After all, there is certainly no interesting topology in ${\bf R^4}$\ to
provide a basis for any other expectation. It was thus of considerable
interest when the existence of counter-examples began to appear around
1982, \cite{f},\cite{d},\cite{g0}.  Our paper
\cite{br} provides a brief survey of this problem and some conjectures
on the possible physical implications of these results.  In this paper,
certain questions raised in \cite{br} are at least partially
answered.\par

Since the existence of non-trivial differentiable structures on
topologically trivial spaces is so strikingly counter-intuitive, it
is important to clarify several issues relating to
differential topology.  Specifically we must distinguish
the case of merely {\em
different} differentiable structures  from
{\em non-diffeomorphic} ones. The former are physically
indistinguishable, but the latter are definitely {\em not} physically
equivalent as
space-time models. These issues are discussed in \cite{br}.  For all
$n\ne 4$ it is possible to show that all smoothness structures on ${\bf
R^n}$ are diffeomorphic. \par However, precisely the opposite is true
for the surprising case, $n=4$, \cite{f},
\cite{d}.  A smooth manifold  homeomorphic to ${\bf R^4}$\ but not
diffeomorphic to it is called ``exotic'' (or ``fake'') and denoted
here by ${\bf R^4_\Theta}$.\footnote{In general, the subscript $\Theta$
will indicate a
non-standard object or process.  So $M\times_\Theta N$ means a smooth
manifold
which is the topological, but not smooth,
cartesian product of the two manifolds. }\ Such a
manifold
 consists of a set of points which
can be globally topologically identified with the ordered set of four
numbers, say $(t,x,y,z)$.  While these may be smooth
coordinates locally over some neighborhood, they cannot be globally
continued
as smooth functions.  Furthermore, in no diffeomorphic image of this
${\bf R^4_\Theta}$\ can the global topological coordinates be extended as
smooth
beyond some compact set.  \par
Also, note that certain ${\bf R^4_\Theta}$\ have the property
that they
contain compact sets which cannot themselves be contained in the interior
of
{\em any} smooth ${\bf S^3}$.  Thus, for some $R_0$, the {\em
topological}
three-sphere,  $t^2+x^2+y^2+z^2=R^2$, cannot be {\em smooth} if $R>R_0$.
This is illustrated in Figure 1. Notice that in Figures 1
through 5 one space dimension has been suppressed, so each point is
actually a z-axis, while in Figure 6 two dimensions are suppressed and
each point is an ${\bf S^2}$.\par
  As interesting as these ${\bf R^4_\Theta}$\ are in their own right, a
technique
developed by Gompf\cite{g1} allows the construction of a large
 topological variety of exotic four-manifolds, some of which would
appear to have considerable potential for physics.
 Gompf's ``end-sum'' process  provides a
straightforward technique for constructing an exotic version, $M$, of any
non-compact four-manifold whose standard version, $M_0$, can be
smoothly embedded
in standard ${\bf R^4}$.  Recall that we want to construct $M$\ which is
homeomorphic
to $M_0$, but not diffeomorphic to it.  First construct a tubular
neighborhood, $T_0$, of a half ray in $M_0$. $T_0$ is thus standard ${\bf
R^4}=
[0,\infty)\times {\bf R^3}$.  Now consider a diffeomorphism, $\phi_0$ of
$T_0$
onto
$N_0=[0,1/2)\times{\bf R^3}$ which is the identity on the ${\bf R^3}$
fibers.  Do
the same thing for some exotic ${\bf R^4_\Theta}$\ with the important
proviso that it
{\em cannot} be smoothly embedded in standard ${\bf R^4}$.  Such
manifolds are known
in infinite abundance \cite{gem}.  Then construct a similar tubular
neighborhood for this ${\bf R^4_\Theta}$, $T_1$, with diffeomorphism,
$\phi_1$, taking
it onto $N_1=[1,1/2)\times{\bf R^3}$.  The desired exotic $M$\ is then
obtained by forming the identification manifold structure
\begin{equation}
M=M_0\cup_{\phi_0}([0,1]\times{\bf R^3})\cup_{\phi_1}\bf
R^4_\Theta\label{es1}\end{equation}
The techniques of forming tubular manifolds and defining
identification manifolds can be found in standard differential
topology texts, such as \cite{bj} or \cite{H}.  \par
Informally, what is being
done is that the tubular neighborhoods are being smoothly glued across
their ``ends'', each ${\bf R^3}$.  The proof that the resulting $M$\ is
indeed  exotic is then easy:  $M$\ contains ${\bf R^4_\Theta}$\ as a
smooth sub-manifold.
If $M$\ were diffeomorphic to $M_0$ then $M$,  and thus ${\bf
R^4_\Theta}$, could be
smoothly embedded in standard ${\bf R^4}$, contradicting the assumption
on
${\bf R^4_\Theta}$.  Finally, it is clear that the constructed $M$\ is
indeed
homeomorphic to the original $M_0$ since all that has been done
topologically is the extension of $T_0$. See figure 2 for a
visualization of this process when $M_0$ is ${\bf R^4}$.  Smoothly
``stuffing'' the upper ${\bf R^4_\Theta}$\ into the tube results in
another
visualization of the new manifold as shown in figure 3.  A natural
doubling of this process leads to figure 4.  Finally, smoothly
spreading out the exotic tube in figure 3 leads to figure 5.
\par
The smoothness properties of the ${\bf R^4_\Theta}$ in figure 4 can be
summarized by saying the global $C^0$ coordinates, $(t,x,y,z)$, are
smooth in the exterior region $[a,\infty){\bf\times S^2\times R^1}$ given
by $x^2+y^2+z^2>a^2$ for some positive constant $a$, while the closure of
the complement of this is clearly an exotic ${\bf B^3\times_\Theta
R^1}$.\footnote{Here the ``exotic'' can be understood as referring to the
product which is continuous but cannot be smooth.  See the discussion
around Lemma 2 below.}  Since the exterior component is standard, a wide
variety, including flat, of Lorentz metrics can be imposed.  Picking only
those for which $\partial/\partial t$ is timelike in this region provides
a natural sense in which the world-tube confining the exotic part is
``spatially localized.''  The smooth continuation of such a metric to the
full metric is then  guaranteed by Lemma 1 and the discussion following
it below.  Thus, we can state\par
\begin{theorem}
There exists smooth manifolds which are homeomorphic but not
diffeomorphic to ${\bf R^4}$ and for which the global topological
coordinates $(t,x,y,z)$ are smooth for $x^2+y^2+z^2\ge a^2>0,$ but not
globally.  Smooth metrics exists for which the boundary of this region is
timelike, so that the exoticness is spatially confined.
\end{theorem}
 \par
We can also use this technique to generate an infinity of non-
diffeomorphic
manifolds, ${\bf R^2\times_\Theta S^2}$, each having the topology of the
Kruskal presentation
of the Schwarzschild metric.  Using the standard Kruskal notation
$\{(u,v,\omega); u^2-v^2<1, \omega\in S^2\}$ constitute global
{\em topological} coordinates, but {\em $(u,v)$ cannot be continued
as smooth functions over the entire range: $u^2-v^2<1$.}  However, by
techniques discussed in \cite{br}, these coordinates can be smooth over
some closed submanifold, say $A$, as illustrated in Figure 6.
Over $A$ then we can solve the vacuum
Einstein equations as usual to get the Kruskal form. However, this metric
cannot be extended over the full manifold, not for any reasons associated
with the development of
singularities in the coordinate expression of the metric, or for any
topological reasons, but simply because {\em the coordinates,
$(u,v,\omega)$, cannot be continued smoothly beyond some proper
subset, $A$,
of the full manifold},
thus establishing
\begin{theorem} On some smooth manifolds
which are topologically
${\bf R^2\times S^2}$, the standard Kruskal metric
cannot be smoothly continued over the full range, $u^2-
v^2<1.$
\end{theorem}\par
However, given any Lorentzian metric on a closed
submanifold, $A$, some smooth continuation of the metric to all of $M$
can be guaranteed to exist under certain conditions.
  For example, we have
\begin{lemmas}
 If $M$ is any smooth connected 4-manifold and $A$ is a closed
submanifold for which
$H^4(M,A; {\bf Z})=0,$ then any smooth, time-orientable
 Lorentz signature metric defined
over $A$
can be smoothly continued to all of $M.$
\end{lemmas}

Proof: This is basically a question of the continuation of cross
sections on fiber bundles.  Standard obstruction theory is usually
done in the continuous category, but it has a natural extension to the
smooth class, \cite{steen}.
First, we note that any time-orientable
 Lorentz metric is decomposable into a Riemannian one, $g$,
plus a non-zero vector field,$v$.  The continuation of $g$ follows from
the fact that the fiber, $Y_S$, of non-degenerate symmetric four by four
matrices is $q$-connected for all $q$.  From standard obstruction theory,
this implies that $g$ can be continued from $A$ to all
of $M$
without any topological restrictions. On the other hand, the fiber of
non-zero vector fields is the three-sphere which is $q$-connected for
all $q<3,$ but certainly not 3-connected ($\pi_3(S^3)={\bf Z}$).  Again
from standard results,
\cite{steen}, any obstruction to a
continuation of $v$ from $A$ to all of $M$
is an element of $H^4(M,A;{\bf Z})$.  Thus, the vanishing of
this group is a sufficient condition for the continuation of $v$,
establishing the Lemma.
\par
In the applications in this paper, $M$ is non-compact, so
$H^4(M;{\bf Z})=0.$ Using the exact cohomology sequence generated by
the
inclusion $A\rightarrow M,$
\begin{equation}
  \cdots\rightarrow H^3(M;{\bf Z})
\rightarrow H^3(A;{\bf Z})\rightarrow H^4(M,A;{\bf Z})
\rightarrow H^4(M;{\bf Z})\rightarrow\cdots\label{exs1}\end{equation}
we see that one way to guarantee the condition of the Lemma is to
have $H^3(A;{\bf Z})=0.$  Another would be to establish that
the map, $H^3(M;{\bf Z})\rightarrow H^3(A;{\bf Z})$ is an epimorphism.
For  example, if $A$ is simply a closed miniature
version of ${\bf R^2\times S^2}$ itself, i.e.,
$A={\bf D^2\times S^2}$, then $H^3(A;{\bf Z})=0$ so
the continuation of a smooth Lorentzian metric is ensured.   Whatever
this metric is, it cannot be the Kruskal one, since otherwise the
manifold would be diffeomorphic to standard ${\bf R^2\times S^2}$.  In
the case discussed in Theorem 1, $A= [a,\infty){\bf\times S^2\times
R^1}$, for which it is also easy to see that $H^3(A;{\bf Z})=0$,
guaranteeing the global smooth metric continuation.
\par
An interesting variation of the situation described in
Figure 6 occurs when $A$  intersects the horizon. Thus it contains a
trapped
surface, so a singularity will inevitably develop from well-known
theorems.  However, if $A$ does not contain a trapped surface
what will happen is
not known.
\par
What is missing from this result, of course, is that the continued metric
satisfy the vacuum Einstein equations and that it be complete in the
Lorentzian sense.  Of course, any smooth Lorentzian metric satisfies
the Einstein equation for some stress-energy tensor, but this tensor
must be shown to be physically acceptable.
 Unfortunately, these issues cannot be resolved
without more explicit information on the global exotic structure than
is presently available.  \par
Another way to study this metric is in terms of
 the original Schwarzschild
$(r,t)$ coordinates, as seen in figure 4.    For this model the
coordinates
$(t,r,\omega)$ are smooth for all of the closed sub-manifold $A$ defined
by
 $r\ge a>2M$ but cannot be continued
as smooth over the entire $M$ or over any diffeomorphic (physically
equivalent) copy.  In this case $A$ is topologically
$[a,\infty)\times{\bf S^2\times R^1}$, so again $H^3(A;{\bf
Z})=0$
and the conditions of lemma 1 are met.  Hence there is some
smooth continuation of any exterior Lorentzian metric in $A$, in
particular,
 the Schwarzschild metric, over the full ${\bf R^4_\Theta}$.  Whatever
this metric
is, it cannot be Schwarzschild since the manifolds are not diffeomorphic.
  An interesting feature of this model is that the manifold is
``asymptotically'' standard in spite of the well known fact that
exotic manifolds are badly behaved ``at infinity''.  However, we note
that this model is asymptotically standard only as
$r\rightarrow\infty,$ but certainly not as $t\rightarrow\infty.$ \par
These models, especially as visualized in figures 3 and 4 are clearly
highly suggestive for investigation of alternative continuation of
exterior solutions into the tube near $r=0.$  We
often discover an {\em exterior}, vacuum solution, and look to continue
it back to some source.  This is a standard problem.  In the stationary
case, we typically have a local, exterior solution to an elliptic
problem, and try to continue it into origin but find we can't as a
vacuum solution unless we have a topology change (e.g., a wormhole),
or unless we add a matter source, changing the  equation.
Now, looking at figures 3 and 4, we are led to consider a third
alternative.\par
Of course, the discussion of stationary solutions involves the idea of
time foliations, which cannot exist globally for these exotic
manifolds, at least not into standard factors.
In fact,
\begin{lemmas}
${\bf R^4_\Theta}$\ cannot be written as a smooth product,
${\bf R^1\times_{smooth} R^3}$. Similarly
 ${\bf R^2\times_\Theta S^2}$\ cannot be written as ${\bf
R^1\times_{smooth}(
R^1\times S^2)}$.
\end{lemmas}
Clearly, if either factor decomposition were smooth, the original
manifold would be standard, since the factors are necessarily
standard from known lower dimensional results, establishing
the lemma.  I am indebted to  Robert Gompf and Duane Randall
 for pointing out to me that because of still open questions it is not
now possible to establish the
 more general result for which the second factor is simply some
smooth three manifold without restriction. Also, note that the question
of factor
decomposition of ${\bf R^4}$ into Whitehead spaces was considered by
McMillan\cite{mc}.\par
Of course, the lack of
a global time foliation of these manifolds means that such models are
 inconsistent with canonical approach to gravity,
quantum theory, etc.  However, it is worth noting that all
experiments yield only local data, so we have no {\em a priori} basis
for excluding such manifolds.\par
These discussions lead naturally to a consideration of what can be
said about Cauchy problems.
Consider then the manifold in figure 5.   The global $(t,x,y,z)$
coordinates are smooth for all $t<0$ but not globally. Now consider,
the Cauchy problem $R_{\alpha\beta}=0$, with flat initial
data on $t=-1$. This is guaranteed to have the complete flat metric
as solution in the standard, ${\bf R^4}$\ case.   However, the similar
problem
{\em cannot} have a complete flat solution for ${\bf R^4_\Theta}$\ since
then the
exponential geodesic map would be a diffeomorphism of ${\bf R^4_\Theta}$\
onto its
tangent space, which is standard ${\bf R^4}$.
This is discussed in \cite{br}.
What must go wrong in the exotic case, of course, is
that $t=-1$ is no longer a Cauchy surface.  However, Lemma 1 can again
be applied here to guarantee the continuation of {\em some} Lorentzian
metric over the full manifold since here $A=(-\infty,-1]\times{\bf R^3}$
so clearly $H^3(A;{\bf Z})=0$.  \par
Finally,  consider the cosmological
model, ${\bf R^1\times_\Theta S^3}$ discussed in \cite{br}.  In this
case, assume
a standard cosmological metric for some time, so here
$A=(-\infty,1]\times{\bf S^3}$.  Clearly,  $H^3(A;{\bf Z})$ does
not vanish in this case, but it can be shown that the inclusion
induced map
$H^3(M;{\bf Z})\rightarrow H^3(A;{\bf Z})$ is onto, so the conditions
of Lemma 1 are met. Thus  some smooth Lorentzian continuation will indeed
exist, leading to some exotic cosmology on ${\bf R^1\times S^3}$.
  \par
I am very grateful to Duane Randall and Robert Gompf for their
invaluable assistance in this work.
\par

\end{document}